\documentclass[aps,pra,twocolumn,superscriptaddress,showpacs,showkeys]{revtex4-1}

\usepackage{graphicx}
\usepackage{amsmath,amssymb}
\begin{document}

%Title of paper
\title{Quantum treatment of two-stage sub-Doppler laser cooling of magnesium atoms}

\author{D.V.~Brazhnikov}
\email{brazhnikov@laser.nsc.ru}

\affiliation{Institute of Laser Physics SB RAS, Novosibirsk 630090, Russia}
\affiliation{Novosibirsk State University, Novosibirsk 630090, Russia}

\author{O.N.~Prudnikov}
\affiliation{Novosibirsk State University, Novosibirsk 630090, Russia}

\author{A.V.~Taichenachev}
\affiliation{Institute of Laser Physics SB RAS, Novosibirsk 630090, Russia}
\affiliation{Novosibirsk State University, Novosibirsk 630090, Russia}

\author{V.I.~Yudin}
\affiliation{Institute of Laser Physics SB RAS, Novosibirsk 630090, Russia}
\affiliation{Novosibirsk State University, Novosibirsk 630090, Russia}
\affiliation{Novosibirsk State Technical University, Novosibirsk 630073, Russia}
%\affiliation{Russian Quantum Center, Skolkovo 143025, Moscow Region, Russia}

\author{A.E.~Bonert}
\affiliation{Institute of Laser Physics SB RAS, Novosibirsk 630090, Russia}

\author{R.Ya.~Il'enkov}
\affiliation{Institute of Laser Physics SB RAS, Novosibirsk 630090, Russia}

\author{A.N.~Goncharov}
\affiliation{Institute of Laser Physics SB RAS, Novosibirsk 630090, Russia}
\affiliation{Novosibirsk State University, Novosibirsk 630090, Russia}
\affiliation{Novosibirsk State Technical University, Novosibirsk 630073, Russia}

\date{\today}

\begin{abstract}
The problem of deep laser cooling of $^{24}$Mg atoms is theoretically studied. We propose two-stage sub-Doppler cooling strategy using electro-dipole transition $3^3$P$_2$$\to$$3^3$D$_3$ ($\lambda\,$$=\,$383.9~nm). The first stage implies exploiting magneto-optical trap with $\sigma^+$ and $\sigma^-$ light beams, while the second one uses a {\it lin}$\perp${\it lin} molasses. We focus on achieving large number of ultracold atoms (T$_{eff}$$<$10$\,\mu$K) in a cold atomic cloud. The calculations have been done out of many widely used approximations and based on quantum treatment with taking full account of recoil effect. Steady-state average kinetic energies and linear momentum distributions of cold atoms are analyzed for various light field intensities and frequency detunings. The results of conducted quantum analysis have revealed noticeable differences from results of semiclassical approach based on the Fokker-Planck equation. At certain conditions the second cooling stage can provide sufficiently lower kinetic energies of atomic cloud as well as increased fraction of ultracold atoms than the first one. We hope that the obtained results can assist overcoming current experimental problems in deep cooling of $^{24}$Mg atoms by means of laser fields. Cold magnesium atoms, being cooled in large number down to several $\mu$K, have certain interest, for example, in quantum metrology.
\end{abstract}

\pacs{37.10.De, 05.10.Gg, 06.30.Ft}

\maketitle

\section{Introduction}

Laser cooling and trapping of neutral atoms plays important role for many directions of modern quantum physics. One of the directions is quantum metrology that experiences galloping progress nowadays. It is aimed on producing of standards of various physical quantities and on carrying out precise measurements with the help of them (e.g., see \cite{Riehle}). At present the most precise measurements are possible for the physical quantities as frequency and time. It is due to success achieved in producing etalons (standards) for these quantities. Contemporary time standard is based on a frequency standard, which defines its stability and accuracy to a considerable degree. At that, frequency etalons can be used not only as a basis for time standards, but also for conducting precise measurements of other physical quantities and constants as, for instance, electrical current and voltage, magnetic field, length, Rydberg and fine-structure constants.

High-accuracy experiments for versatile examination of relativistic and quantum theories have become feasible owing to modern frequency standards. Among practical applications of time and frequency standards the broadband communication networks, navigational and global positioning systems should be mentioned especially. Many laboratories in word-known scientific centers do research in the field of frequency standards. One of the latest trends in this field is connected with the concept of intercity or even international quantum clock network that could combine time etalons from various laboratories and countries into one system \cite{Network1,Network2,Network3,Network4,Network5}.

There are two main directions of primary frequency standards development: the technology of a single ion confined in a electro-quadrupole trap and the second one based on many neutral atoms trapped in an optical lattice (e.g., see \cite{KatoriMetrology,Lemonde}). The latter direction is much newer than the former and it undergoes intense progress. The idea of neutral atoms trapping in a periodic light potential is not new and it was actively studied in 1970s (see monograph \cite{Kazantsev} and citations in it). As for metrological purposes this idea has experienced the second birth in the beginning of XXI century after noticeable progress in technique and methods of laser cooling of atoms, development of the ``magic''-wavelength concept \cite{Magic1,Magic2}, and also experimental and theoretical success in the field of spectroscopy of forbidden atomic transitions \cite{SrKatoriMagic,YbTransition,Hong,BizeSpectroscopy,YudinRamsey}. At present, stability of optical-lattice-based frequency standards is on the same level with single-ion standards and in some cases even better. The state-of-the-art prototypes reached instability and uncertainty on the relative levels of $10^{-17}$$-$$10^{-18}$ \cite{NatureKatori,SrPTB,NatureYe,Katori2015}.

\begin{figure}[t]
\centerline{\scalebox{0.6}{\includegraphics{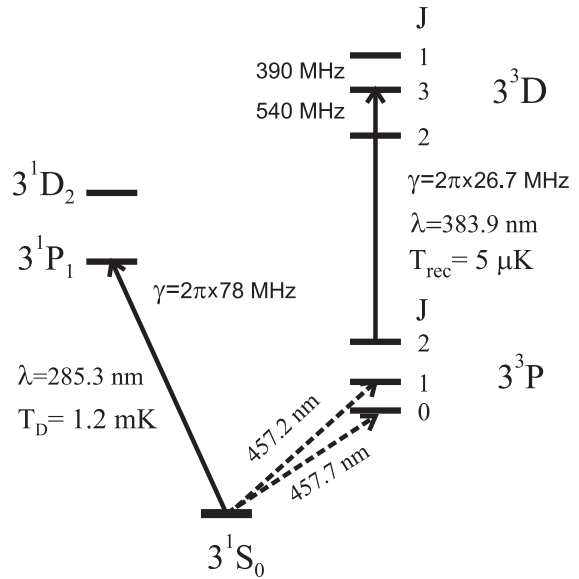}}}
\caption{Partial energy diagram of $^{24}$Mg atom. Solid lines denote the cooling transitions with corresponding temperature limits, while dashed lines denote possible ``clock'' transitions, which can be used for laser stabilizing.}\label{fig1}
\end{figure}

Ones of the main candidates for producing the new-generation frequency standards are alkaline earth and alkaline-earth-like atoms: Yb (for instance, see \cite{YudinMagnetic,YbMagic,YbNIST}), Ca \cite{CaMagic}, Sr \cite{SrPTB,SrKatori,NatureYe}, Hg \cite{HgMagic} and Mg \cite{MgErtmer,MgGoncharov}. These atoms are the most appropriate because of narrow spectroscopic lines due to forbidden optical transitions from the ground state $^1$S$_0$ to the lowest excited triplet state $^3$P$_{0,1,2}$ (see Fig.~\ref{fig1}). Moreover, one more key circumstance consists in the existence of so-called ``magic'' wavelength for these transitions. Under the magic-wavelength optical field the linear (in the intensity) light shift is canceled.  Also, one of the last tendency in this field is connected with spectroscopy of transition $^1$S$_0$$\to$$^3$P$_0$ in even isotopes (with zero nucleus spin), which is highly forbidden. Frequency of this transition is immune to many frequency-shift effects, therefore this transition can be exploited as a good ``clock'' transition. In spite of the transition is highly forbidden, it has already been observed by means of magnetic-field-induced spectroscopy \cite{YudinMagnetic2} in $^{174}$Yb \cite{YudinMagnetic}, $^{88}$Sr \cite{SrKatori,SrKatoriNature,SterrIEEE,SterrPRL,SrClockTino} and $^{24}$Mg \cite{MgDeep1}.

To date, atoms of the first four elements (Yb, Ca, Sr and Hg) can be effectively cooled by the laser methods down to ultralow temperatures, approaching to the recoil energy limit, what is required for effective loading an optical lattice ($\sim$ 1--10~$\mu$K). Besides, sub-recoil temperatures can be obtained by using evaporative cooling technique, getting Bose-Einstein condensation \cite{BECYb,BECCa,BECSr}. Unfortunately, researchers have not been able to reach the same great success with Mg atoms. In particular, neither two-photon laser cooling \cite{Twophoton} nor laser quenching \cite{Quenching} methods have appeared to be ineffective in the case with magnesium. The minimum temperature of a magnesium cloud that has been obtained by laser cooling method is about 500~$\mu$K, what is rather far from desirable range of values, in particular, from the recoil temperature (3--10~$\mu$K, depending on an atomic transition).

At the same time, magnesium atom has some advantages with respect to the other candidates for the frequency standard. Thus, black-body radiation (BBR) shift is one of the main limiting factors for accuracy and stability of quantum frequency standard (e.g., see \cite{NatureYe,Network1,Lemonde}). BBR shift of the clock transition $3^1$S$_0$$\to$$3^3$P$_0$ for magnesium is much smaller than for Yb, Ca, Sr and just a little bit higher than for a mercury atom (see Tab.1). However, from an experimental viewpoint Mg has some advantages against Hg. In particular, mercury atom requires noticeably smaller wavelengths for laser spectroscopy, cooling and trapping than Mg. For instance, effective trapping of cold atoms in a nondissipative optical lattice and producing Lamb-Dicke regime need for an optical potential depth at the level of 50--300 in the recoil energy units (e.g., see \cite{NatureYe,Lemonde}). It means that a highly intensive laser field at the magic wavelength $\lambda_m$ should be applied. But, as it can be seen from Table 1, $\lambda_m$(Mg)$\approx$$\,468$~nm and $\lambda_m$(Hg)$\approx$$\,363$~nm. Therefore, from an experimental viewpoint, producing deep optical potential for mercury atom is more difficult problem than for magnesium one due to the much smaller wavelength. Besides of relatively small BBR shift, magnesium atom has one more advantage with respect to Ca and Sr. It consists in absence of optical pumping of atoms on the non-resonant level $3^1$D$_2$ during the realization of laser precooling with the help of strong dipole transition $3^1$S$_0$$\to$$3^1$P$_1$ (see Fig.~\ref{fig1}) to reach the temperature down to a few millikelvins \cite{MgErtmer,MgGoncharov,DopplerMg1,DopplerMg2}.

%\begin{table}[t]\label{tab1}
%\caption{Table}
%\end{table}

\begin{table}[t]\label{tab1}
\caption{Data for several atomic elements relevant for new-generation frequency standards: $\lambda_{cl}$ is a wavelength of the clock transition $3^1$S$_0$$\to$$3^3$P$_0$ and $\lambda_m$ is its magic wavelength, BBR frequency shifts are indicated with respect to absolute frequencies of clock transition ($\lambda_{cl}$ is taken from NIST Atomic Spectra Database\footnote{http://physics.nist.gov/asd}).}
\begin{ruledtabular}
\begin{tabular}{c|c|c|c}
Atom & $\lambda_{cl}$ & $\lambda_m$ & BBR shift \\ \hline
Sr & 698.5 & 813.5 \cite{SrKatoriMagic} & $-5.5$$\times$$10^{-15}$ \cite{PortsevBBR}  \\ \hline
Yb & 578.4 & 759.4 \cite{YudinMagnetic,YbMagic} & $-2.6$$\times$$10^{-15}$ \cite{PortsevBBR}\\ \hline
Ca & 659.7 & 735.5 \cite{CaMagic} & $-2.6$$\times$$10^{-15}$ \cite{PortsevBBR} \\ \hline
{\bf Mg} & {\bf 457.7} & {\bf $\approx$ 468} \cite{MgMagic} & ${\bf -3.9\times 10^{-16} }$ \cite{PortsevBBR} \\ \hline
Hg & 265.6 & 362.6 \cite{HgMagic} & $-2.4$$\times$$10^{-16}$ \cite{BBRHg1,BBRHg2}\\
\end{tabular}
\end{ruledtabular}
\end{table}

Recent experiments \cite{MgDeep1,MgDeep2} showed some progress in cooling of $^{24}$Mg atoms. The atoms were cooled down to the record temperature equaled to 1.3~$\mu$K and confined in an optical lattice. However, the final number of atoms was of the order of $10^4$, what was about 0.01\,\% from the initial number of atoms in a magneto-optical trap (MOT), involved the cyclic triplet dipole transition $3^3$P$_2$$\to$$3^3$D$_3$. That great loss in atomic number was due to the fact that velocity selection technique, having similarities with evaporative cooling, was used for reaching such ultralow temperature, but the laser cooling method in MOT, unfortunately, showed the cloud temperature equaled to only about 1~mK. That result noticeably yielded to successful results with the other atomic elements (Ca, Sr, Yb, Hg). At the same time, we believe that laser cooling strategy for magnesium atoms can be proper tuned for getting much better result of laser-cooling temperature as well as number of atoms trapped.

Therefore, we can state that problem of deep cooling of magnesium atoms by means of laser radiation is still unsolved. Moreover, increasing of ultracold atomic number has principal importance for many applications of cold atoms. For instance, authors of the paper \cite{BECSr2} managed to obtain Bose-Einstein condensation composed of $\sim$$10^7$ strontium atoms. Besides, frequency-standard stability depends on an atomic number in an optical lattice and it increases with the number increases \cite{Stability,Riehle}. All these things considered, we can conclude that it is important to solve the problem of deep laser cooling of magnesium atoms (down to T$\sim$1--10~$\mu$K) as well as to provide much larger number of ultracold atoms in a lattice.

\section{Laser cooling in MOT: Semiclassical approximation}\label{Paragraph2}
Laser cooling of neutral atoms in a magneto-optical trap is one of the main cooling methods. At first time the laser field composed of six beams with orthogonal circular polarizations ($\sigma^+\sigma^-$ configuration) was suggested in \cite{DalibardSigma} as effective way for simultaneous cooling and trapping of atoms. Narrow spectral lines allow laser cooling of various atoms down to a few tens and units of microkelvin, and even lower. In particular, narrow intercombination transition $4^1$S$_0$$\to$$4^3$P$_1$ in $^{40}$Ca ($\gamma$$\approx$$2\pi$$\times$$400$\,Hz) provides temperatures around 4--6~$\mu$K \cite{CaCooling1,CaCooling2} just by the help of Doppler cooling process. There were good results with intercombination transitions also for the other elements: Sr \cite{SrCooling}, Yb \cite{YbCooling,YudinMagnetic} and Hg \cite{BBRHg1}, for even as well as for odd isotopes. In certain aspects even isotopes, having a zero nuclear spin, are more attractive for frequency standards of new generation. However, as it has been already noted in the Introduction, still there are not satisfactory results of cooling $^{24}$Mg atoms by means of laser radiation in contrast to the other elements.

The first our attempt to solve the problem with deep laser cooling of magnesium was undertook in the recent work \cite{MgSemiclassics}, where the detailed theoretical study of magnesium kinetics in 1D MOT using the dipole transition $3^3$P$_2$$\to$$3^3$D$_3$ was conducted. The theory was based on the semiclassical approach \cite{Kazantsev,Minogin}, based on the well-known assumptions:

\begin{equation}\label{eq1}
\omega_{rec}\ll \text{min}\{\gamma,\gamma S\}\,
\end{equation}

\noindent and

\begin{equation}\label{eq2}
\Delta p\gg\hbar k\,.
\end{equation}

\noindent Here $\omega_{rec}$$=$$\hbar$$k^2$$/2M$ is the recoil frequency, $M$ is mass of an atom, $k$$=$$2$$\pi/$$\lambda$ is wave number. The saturation parameter $S$ is defined as

\begin{equation}\label{saturation}
S=\frac{R^2}{(\gamma/2)^2+\delta^2}\,,
\end{equation}

\noindent where $\gamma$ is the spontaneous relaxation rate of excited state, $\delta$$=$$\omega$$-$$\omega_0$ is the detuning of laser radiation frequency $\omega$ from the transition frequency $\omega_0$, and $R$ is the Rabi frequency.

Condition (\ref{eq1}) implies that recoil frequency must be rather small in comparison with a typical rate of establishment of steady state among atomic internal degrees of freedom. In particular, in the case of an atom without any degeneracy of the ground state this rate is defined by $\gamma$. If there is a degenerate ground state and optical pumping can occurs, this rate is defined by $\gamma$ or the pumping rate $\gamma$$S$, depending on what is smaller. The second semiclassical requirement (\ref{eq2}) implies that typical width of stationary linear momentum distribution $f(p)$ must be much larger than the recoil momentum from emmision/absorption of a photon.

Doppler limit for temperature of laser cooling $T_D$, which can be achieved at the frequency detuning $\delta$$=$$-\gamma$$/2$, can be figured out from equation for minimum kinetic energy in one-dimensional case:

\begin{equation}\label{DopplerLimit}
E^{min}_{kin}=\frac{1}{2}k_B T_D = \frac{7}{40}\hbar\gamma.
\end{equation}

\noindent Strictly speaking, this equation is valid for transition $J_g$$=$$0$$\to$$J_e$$=$$1$. It was found in \cite{CastinDoppler} under $\sigma^+$$\sigma^-$ configuration (also see \cite{KazantsevLimit}). If we use this formula for getting estimate of $T_D$ in the case of transition $3^3$P$_2$$\to$$3^3$D$_3$ ($\gamma$$\approx$$2\pi$$26.7\,$MHz), we immediately find $T_D$$\approx\,$$425\,$$\mu$K. For effective trapping of atoms with such relatively high temperature the large intensity of cw optical lattice field at the level of tens of  MW/cm$^2$ is required, what is hardly feasible in an experiment. Therefore, much lower temperature of atomic cloud is needed. At the same time, since the transition considered has degenerate energy levels, one can anticipate activation of so-called sub-Doppler mechanism during the laser cooling in MOT under the polarization-gradient field. In principal, this process would overcome the Doppler limit (\ref{DopplerLimit}) and show much lower temperature than in the case of $J_g$$=$$0$$\to$$J_e$$=$$1$.

Semiclassical approach is based on kinetic equation of Fokker-Planck type on the Wigner distribution function in phase space $f(z,p)$. That equation can be acquired by reducing of exact quantum kinetic equation on the density matrix in the series on small parameter $\hbar k$$/\Delta p$$\ll$$1$ until second-order terms. This procedure is well-known and it has been done by many authors (e.g., see \cite{MinoginKinetics,Cook,DalibardKinetics,Juha91}). Eventually the following equation can be obtained:

\begin{equation}\label{Semiclassics}
\frac{p}{M}\,\frac{\partial}{\partial z}f(z,p) = \Bigl[-\frac{\partial}{\partial p}F(z,p)+\frac{\partial^2}{\partial p^2}D(z,p)\Bigr]\,f(z,p).
\end{equation}

\noindent Here $F(z,p)$ is laser-field force on an atom, $D(z,p)$ is diffusion of an atom in the light field. This equation must be completed with normalizing condition that in one-dimensional periodic laser field has the form:

\begin{eqnarray}
\frac{1}{\lambda}\int\limits_{-\lambda/2}^{+\lambda/2}dz\int\limits_{-\infty}^{+\infty}f(p,z)dp=1\,.\nonumber
\end{eqnarray}

\noindent One-dimensional $\sigma^+\sigma^-$ laser-field configuration allows significant simplifying of (\ref{Semiclassics}), at that the dependence $f$ on $z$ vanishes (see section \ref{FirstStage}).

Our semiclassical calculations \cite{MgSemiclassics} have been done beyond many widely used approximations (for instance, slow atoms and weak field approximations). As it has been shown the minimum kinetic energy achievable in MOT is close to 30$\times$$E_{rec}$, where $E_{rec}$$=$$\hbar\omega_{rec}$ is the recoil energy. The effective temperature, which can be associated to this value, $T_{eff}$$\approx$$150$$\,\mu$K. It is approximately three times lower than the estimate of Doppler limit $T_D$$\approx$$\,425$$\,\mu$K, but, unfortunately, it is still very far from desirable range of values and, in particular, the recoil temperature $T_{rec}$$=$$5$$\,\mu$K.

\begin{figure}[t]
\centerline{\scalebox{0.8}{\includegraphics{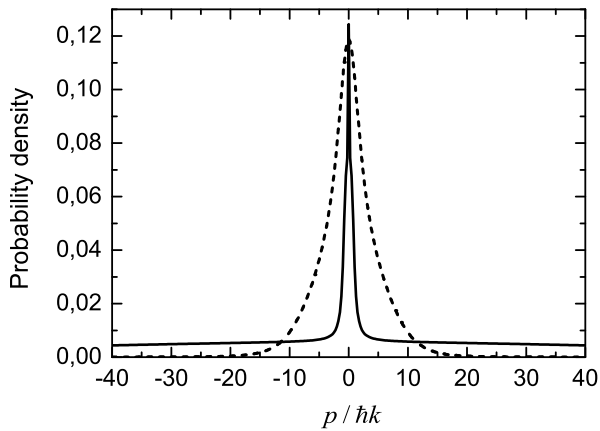}}}
\caption{Momentum distributions of magnesium atoms at $\delta$$=$$-5$$\gamma$$\approx$$-2\pi$$\times$$130\,$MHz, $I$$=$$20\,$mW/cm$^2$ (solid) and $I$$=$$470\,$mW/cm$^2$ (dashed).} \label{fig2}
\end{figure}

Let us consider a question on validity of semiclassical approach to the magnesium problem. Indeed, as it will be shown further on the basis of semiclassical treatment, the optimal parameters of cooling field for transition $3^3$P$_2$$\to$$3^3$D$_3$ can be chosen as $\delta$$=$$-2\pi$$\times$$130$~MHz and $I$$=$$500\,$mW/cm$^2$. The corresponding saturation parameter $S$$\approx$$4$$\times$$10^{-2}$. In spite of such low saturation, the first semiclassical requirement (\ref{eq1}) is still satisfied, because $\omega_{rec}$$=$$2$$\times$$10^{-3}$$\,\gamma$. At the same time, typical momentum distribution width $\Delta p$ may not satisfied the second semiclassical condition (\ref{eq2}). Indeed, in general case the distribution can have complex shape. In particular, Fig.~\ref{fig2} shows two examples of momentum distributions for different values of light field intensity. The distribution acquires two-peaked profile at low intensity $I$$=$$20\,$mW/cm$^2$: there is the high-contrast spike on top of the wide background. This background conditionally describes ``hot'' fraction of atoms in a cloud with effective temperature $T_{eff}$$\sim$$\,1$$-$$10\,$~mK, while the spike corresponds to ultracold fraction with $T_{eff}$$\sim$$\,1\,\mu$K. Similar distributions were observed earlier (e.g., see \cite{CsDistribution} with semiclassical low-saturation-limited calculations of Sisyphus cooling of Cs atoms with the help of transition $F_g$$=$$4$$\to$$F_e$$=$$5$ or the quantum-treatment calculations for atomic W-type scheme in \cite{Dalibard}). In our case the narrow-spike width is about $\hbar k$ and the requirement $\Delta p$$\gg$$\hbar k$ is not satisfied at all. With increasing the intensity ($I$$=$$470\,$mW/cm$^2$) the two-peaked shape disappears. However, the distribution as a whole is still sufficiently narrow and the second requirement (\ref{eq2}) is satisfied with a good margin. Also it should be noted that the condition (\ref{eq2}), as a matter of fact, depends on the value of total angular momentum $F_g$. In other words, at the same saturation parameter $S$ the requirement (\ref{eq2}) can be valid for small $F_g$ and getting not valid with its increasing.

Basing on the aforesaid, we can conclude that more precise theoretical treatment is needed in the case with magnesium for adequate description of kinetics of ultracold atoms. This treatment can be based on the density matrix formalism with full account for the recoil effect (e.g., see \cite{Kazantsev,MinoginKinetics,Prudnikov}). Moreover, as it will be seen in the next section, the quantum-treatment results noticeably differs from the semiclassical ones, based on the equation (\ref{Semiclassics}). That difference, in particular, gave us an idea for exploiting the second stage of sub-Doppler laser cooling for getting the desirable results.

\section{Full account for the recoil effect}
Let us consider the problem of laser cooling of magnesium atoms out of semiclassical approximation limit as well as some other widely used approximations (weak-saturation limit, secular approximation, etc.).

\subsection{Problem statement}
We assume the laser field to be one-dimensional, composed of two plane monochromatic counterpropagating light waves with equal frequencies and amplitudes (the quantization axis $z$ is collinear to the wave vectors):

\begin{eqnarray}\label{Field}
{\bf E}(z,t)&&=E_0{\bf e}_1 e^{-i(\omega t-kz)}+E_0{\bf e}_2 e^{-i(\omega t+kz)}+c.c.=\nonumber\\
&&=E_0{\bf e}(z)\,e^{-i\omega t}+c.c.,
\end{eqnarray}

\noindent where ${\bf e}_{1,2}$ are the unit complex vectors of waves' polarizations, while ${\bf e}(z)$ is the following complex vector

\begin{eqnarray}\label{FieldPolarization1}
{\bf e}(z)={\bf e}_1 e^{ikz}+{\bf e}_2 e^{-ikz}\,.
\end{eqnarray}

\noindent Nonzero components of the vectors ${\bf e}_{1,2}$ in the spherical basis are

\begin{eqnarray}\label{Basis}
&&e_1^{-1}=-\sin(\varepsilon_1-\pi/4),\,\,e_1^{+1}=-\cos(\varepsilon_1-\pi/4),\nonumber\\
&&e_2^{-1}=-\sin(\varepsilon_2-\pi/4)\,e^{i\varphi},\nonumber\\
&&e_2^{+1}=-\cos(\varepsilon_2-\pi/4)\,e^{-i\varphi}.
\end{eqnarray}

\noindent Here $\varepsilon_{1,2}$ are the ellipticity parameters (in particular, $\varepsilon$$=$$\pm\pi/4$ corresponds ro righ- or left-circular polarized wave, $\varepsilon$$=$$0$ is for linear polarization), $\varphi$ is the angle between main axes of polarization ellipses. For instance, the case with $\varepsilon_{1,2}$$=$$0$ and $\varphi$$=$$\pi/2$ corresponds to {\it lin}$\perp${\it lin} field configuration.

Here quantum treatment of atomic kinetics under the laser field (\ref{Field}) is based on the equation on single-atom density matrix in coordinate two-point representation that has the form (e.g., see \cite{Minogin,Kazantsev,Prudnikov}):

\begin{equation}\label{Density1}
\frac{\partial\widehat{\rho}(z_1,z_2,t)}{\partial t} = -\frac{i}{\hbar}\Bigl[\widehat{H}(z_1,t)\,\widehat{\rho}-\widehat{\rho}\,\widehat{H}(z_2,t)\Bigr]+\widehat{\Gamma}\bigl\{\widehat{\rho}\bigr\},
\end{equation}

\noindent with the Hamiltonian

\begin{equation}\label{Hamiltonian}
\widehat{H}(z_i,t)=(\widehat{p}_i^2/2M)+\widehat{H}_0+\widehat{V}(z_i,t)\,.
\end{equation}

\noindent The first term in the Hamiltonian is the operator of kinetic energy of an atom ($\widehat{p}_i$ is the linear momentum operator), $\widehat{H}_0$ describes intratomic degrees of freedom, operator $\widehat{V}$ corresponds to the atom-field dipole interaction, and the linear operator functional $\widehat{\Gamma}$$\{\dots\}$ is respective for relaxation processes in an atom. Let us introduce the projection operator onto the excited atom state:

\begin{equation}\label{Projections}
\widehat{P}^e=\sum_{m_e}\mid F_e,m_e\rangle\langle F_e,m_e\mid\,\,,
\end{equation}

\noindent and the Wigner vector operator $\widehat{{\bf T}}$, whose spherical components are:

\begin{equation}\label{T}
\widehat{T}_\sigma=\sum_{m_e,m_g}C^{F_e,m_e}_{F_g,m_g;1\sigma}\mid
F_e,m_e\rangle \langle F_g,m_g\mid\,,
\end{equation}

\noindent with $\sigma$$=$$0,\,\pm1$ and $C^{F_e,m_e}_{F_g,m_g;1\sigma}$ the Clebsch-Gordan coefficients (e.g., see \cite{Varshalovich}). Then the terms $\widehat{H}_0$ and $\widehat{V}$ from (\ref{Hamiltonian}) in the resonant approximation can be written as

\begin{equation}\label{H0}
\widehat{H}_0=-\hbar\delta\widehat{P}^e\,
\end{equation}

\noindent and

\begin{equation}\label{V}
\widehat{V}(z_i)=-\hbar R \,\widehat{{\bf T}}\cdot{\bf e}(z_i)\,+h.c.=-\hbar R \,\widehat{V}^{eg}(z_i)\,+h.c.
\end{equation}

\noindent Here $R$$=$$E_0d/\hbar$ is the Rabi frequency (with $d$ the reduced matrix element of dipole operator of an atom), $\widehat{V}^{eg}(z_i)$$=$$\widehat{{\bf T}}$$\cdot$$\,{\bf e}(z_i)$ is the dimensionless operator of atom-field interaction, depending on the coordinate in general case, $h.c.$ means Hermitian-conjugate term.

Introduce the new coordinates:

\begin{equation}\label{Coordinates}
z=k(z_1+z_2)/2, \quad q=k(z_1-z_2)\,,
\end{equation}

\noindent in which the spontaneous relaxation operator from (\ref{Density1}) acquires the form:

\begin{eqnarray}\label{Gamma}
\widehat{\Gamma}=-\frac{\gamma}{2}\Bigl(\widehat{P}^e\widehat{\rho}+\widehat{\rho}\widehat{P}^e\Bigr)
+\gamma\sum_{\sigma=0,\pm1}\zeta_\sigma(q)\widehat{T}_\sigma^{\dag}\rho\widehat{T}_\sigma\,,
\end{eqnarray}

\noindent with

\begin{eqnarray}\label{Zeta}
&&\zeta_{\pm1}=\frac{3}{2}\Bigl(\frac{\sin(q)}{q}-\frac{\sin(q)}{q^3}+\frac{\cos(q)}{q^2}\Bigr)\,,\nonumber\\
&&\zeta_0=3\Bigl(\frac{\sin(q)}{q^3}-\frac{\cos(q)}{q^2}\Bigr)\,.
\end{eqnarray}

\noindent Note that in the absence of recoil effect, i.e. in the limit $q\to0$, we have $\zeta_\sigma=1$.

The density matrix can be divided into four matrix blocks:

\begin{equation}\label{DansityMatrix}
\widehat{\rho}=
\left(
\begin{array}{cc}
\widehat{\rho}^{gg}& \widehat{\rho}^{ge}\\
\widehat{\rho}^{eg}& \widehat{\rho}^{ee}
\end{array}
\right)\,.
\end{equation}

\noindent Matrix blocks $\widehat{\rho}^{gg}$ and $\widehat{\rho}^{ee}$ describes populations of the ground and the excited states as well as low-frequency (Zeeman) coherences. Blocks $\widehat{\rho}^{ge}$ and $\widehat{\rho}^{eg}$ are responsible for optical coherences. For the new coordinates (\ref{Coordinates}) and using all introduced notations the new equations on the density matrix blocks can be easily acquired from (\ref{Density1}). So, in the steady state we have

\begin{eqnarray}\label{DensityBlocks}
&&-2i\omega_r\frac{\partial^2}{\partial q \partial z}\widehat{\rho}^{gg}(z,q) =\gamma\sum_{\sigma=0,\pm1}\zeta_\sigma(q)\widehat{T}_\sigma^{\dag}\rho\widehat{T}_\sigma+\nonumber\\
&&\qquad\quad +iR\Bigl[  \widehat{V}^{eg\,\dag}(z+\frac{q}{2}) \widehat{\rho}^{eg}  - \widehat{\rho}^{ge} \widehat{V}^{eg}(z-\frac{q}{2})  \Bigr]\,,\\
&&\Bigl(\gamma-2i\omega_r\frac{\partial^2}{\partial q \partial z}\Bigr)\widehat{\rho}^{ee}(z,q)=\nonumber\\
&&\qquad =iR\Bigl[  \widehat{V}^{eg}(z+\frac{q}{2}) \widehat{\rho}^{ge}  - \widehat{\rho}^{eg} \widehat{V}^{eg\,\dag}(z-\frac{q}{2})  \Bigr]\,,\\
&&\Bigl(\frac{\gamma}{2}+i\delta-2i\omega_r\frac{\partial^2}{\partial q \partial z}\Bigr)\widehat{\rho}^{ge}(z,q)=\nonumber\\
&&\qquad =iR\Bigl[  \widehat{V}^{eg\,\dag}(z+\frac{q}{2}) \widehat{\rho}^{ee}  - \widehat{\rho}^{gg} \widehat{V}^{eg\,\dag}(z-\frac{q}{2})  \Bigr]\,,\\
&&\Bigl(\frac{\gamma}{2}-i\delta-2i\omega_r\frac{\partial^2}{\partial q \partial z}\Bigr)\widehat{\rho}^{eg}(z,q)=\nonumber\\
&&\qquad =iR\Bigl[  \widehat{V}^{eg}(z+\frac{q}{2}) \widehat{\rho}^{gg}  - \widehat{\rho}^{ee} \widehat{V}^{eg}(z-\frac{q}{2})  \Bigr]\,.\label{DensityBlocks4}
\end{eqnarray}

\noindent These equations compose a basis for further theoretical analysis. For instance, probability density of atoms in the momentum space can be found from the formula:

\begin{eqnarray}\label{Distribution}
f(p)=\frac{1}{(2\pi)^2}\int\limits_{-\infty}^{+\infty}dq\int\limits_{-\pi}^{\pi}dz\,\text{Tr}\bigl\{\widehat{\rho}(z,q)\bigr\}e^{-ipq}\,.
\end{eqnarray}

\noindent Here the linear momentum of an atom evaluated in the recoil momentum units $\hbar k$ and $\text{Tr}[\dots]$ denotes trace operation. Momentum distribution $f(p)$ must be normalized:

\begin{eqnarray}\label{Norm1}
\int\limits_{-\infty}^{+\infty}f(p)dp=1\,.
\end{eqnarray}

\noindent This means that the set of equations (\ref{DensityBlocks})-(\ref{DensityBlocks4}) must be supplemented with the condition:

\begin{equation}\label{Norm2}
\frac{1}{2\pi}\int\limits_{-\pi}^{\pi}\text{Tr}\bigl\{\widehat{\rho}(z,q=0)\bigr\}dz=1\,.
\end{equation}

\noindent Average kinetic energy of an atom in the recoil energy units can be evaluated, for instance, with the help of the following formula:

\begin{eqnarray}\label{Ekin}
E_k=\int\limits_{-\infty}^{+\infty}p^2f(p)dp\,.
\end{eqnarray}

\subsection{The first stage: Cooling in MOT}\label{FirstStage}
As a rule, in magneto-optical trap atoms are localized in weak-magnetic-field region (in the vicinity of trap's center). Therefore, magnetic field does not affect significantly on the temperature of a cloud and we omit it here from our analysis. In other words, we consider kinetics of atoms in 1D laser field, composed of two counterpropagating beams with orthogonal circular polarizations ($\sigma^+$$\sigma^-$ configuration). Then in equation (\ref{Basis}) we can take $\varepsilon_1$$=$$\pi/4$, $\varepsilon_2$$=$$-\pi/4$, $\varphi$$=$$0$ and polarization vector of total light field in spherical basis takes the form:

\begin{eqnarray}\label{FieldPolarizationLin}
{\bf e}(z_i)={\bf e}_{-1}e^{-ikz_i}-{\bf e}_{+1}e^{ikz_i}\,.
\end{eqnarray}

\noindent This form corresponds to the laser field with linear polarization, which rotates by an angle $\alpha$$=$$-kz_i$ during propagation along $z$-axis. At that the dimensionless operator $\widehat{V}^{eg}$ from (\ref{V}) is

\begin{eqnarray}\label{VegMOT}
\widehat{V}^{eg}(z_i)=\widehat{T}_{-1}e^{-ikz_i}-\widehat{T}_{+1}e^{ikz_i}\,.
\end{eqnarray}

\begin{figure}[t]
\centerline{\scalebox{0.6}{\includegraphics{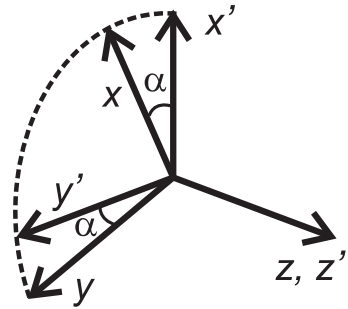}}}
\caption{Transformation the old coordinate frame $K$ to the new one $K'$.}\label{fig3}
\end{figure}

\begin{figure*}[t]
\centerline{\scalebox{0.9}{\includegraphics{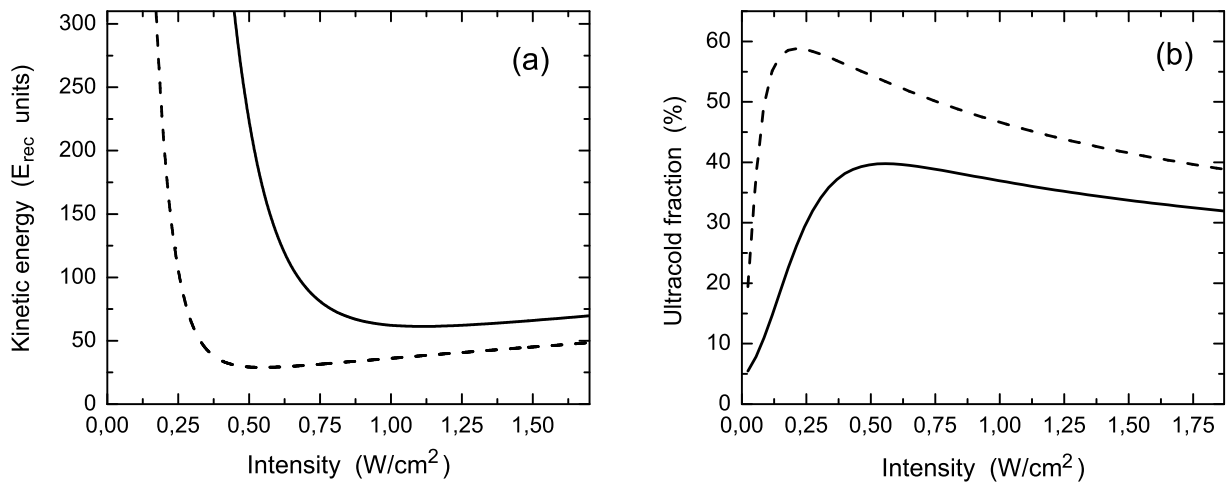}}}
\caption{Comparison the results of semiclassical (dashed) and quantum (solid) treatments at $\delta$$=$$-5$$\gamma$$\approx$$-2\pi$$\times$$130\,$MHz. (a) Average kinetic energy of an atom as the function of light field intensity, (b) Ultracold fraction of atoms in a cloud.}\label{fig4}
\end{figure*}

\noindent The considered field configuration has some unique features. First of all, the field has homogeneous intensity (it does not depend on $z$-coordinate). And the second, the field polarization also can be made homogeneous (e.g., see \cite{Dalibard,YudinLasPhys}). Indeed, let us pass to the new coordinate system $K'$, in which $z'$-axis coincides with $z$-axis in the old $K$-system, while the axes $x'$ and $y'$ rotate around $z$-axis by the angle $\alpha$$=$$-z$$=$$-k$$(z_1$$+$$z_2$$)/2$ (see Fig.~\ref{fig3}). In the $K'$-system the linearly polarized total-field vector does not rotate anymore (without loss of generality it can be considered to be directed along $x'$-axis). Then in the new system the interaction operator $\widehat{V}^{eg}$ from (\ref{V}) does not depend on the coordinate $z_i$:

\begin{eqnarray}\label{VegMOTNew2}
&&\widehat{V}^{eg}(z_1)\,\Longrightarrow\,\widehat{V}_1(q)=\widehat{T}_{-1}e^{-iq/2}-\widehat{T}_{+1}e^{iq/2}\,,\\
&&\widehat{V}^{eg}(z_2)\,\Longrightarrow\,\widehat{V}_2(q)=\widehat{V}_1^*(q)\,.
\end{eqnarray}

\noindent Since the relaxation operator $\widehat{\Gamma}$ from (\ref{Gamma}) also does not depend on $z$, the density matrix in the new coordinate system is not the function of $z$ and it depends only on $q$-coordinate. This circumstance significantly simplifies numerical evaluations of the density matrix equations.

The operator of rotation $\widehat{D}({\bf n},\alpha)$ can be exploited for getting the equations on density matrix in the new basis (e.g., see \cite{Varshalovich}). Here the unit vector ${\bf n}$ defines a rotation axis, while $\alpha$ is a rotation angle. In our case it is natural to coincide ${\bf n}$ with quantization axis $z$ and take $\alpha$$=$$-z$. Then action of the rotation operator on a wave function $\mid F_a,m_a,z_i\rangle$ reduces to a simple multiplication by $e^{im\alpha}$, i.e.

\begin{eqnarray}\label{Rotation}
\widehat{D}({\bf n},\alpha)\mid F_a,m_a,z_i\rangle=e^{im_a\alpha}\mid F_a,m_a,z_i\rangle\,,
\end{eqnarray}

\noindent with $(a$$=$$e,g)$. In the system $K'$ the set of equations (\ref{DensityBlocks})-(\ref{DensityBlocks4}) take the form:

\begin{eqnarray}\label{DensityRotated2}
2\omega_r\frac{\partial}{\partial q}&&\Bigl[\widehat{F}_z,\widehat{\rho}(q)\Bigr]=\widehat{\Gamma}\{\widehat{\rho}(q)\}+i\delta\Bigl[\widehat{P}^e,\widehat{\rho}\Bigr]+\nonumber\\
&&+\,iR\,\Bigl[\widehat{V}_1(q)\widehat{\rho}-\widehat{\rho}\widehat{V}_2(q)\Bigr]\,.
\end{eqnarray}

\noindent At that the normalizing condition (\ref{Norm2}) becomes rather simple:

\begin{equation}\label{Norm3}
\text{Tr}\bigl\{\widehat{\rho}(q=0)\bigr\}=1\,.
\end{equation}

\begin{figure}[t]
\centerline{\scalebox{0.7}{\includegraphics{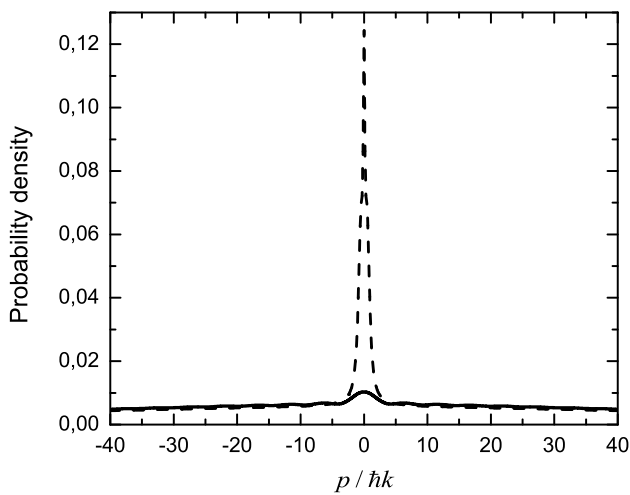}}}
\caption{Momentum distributions of magnesium atoms: comparison of semiclassical (dashed) and quantum (solid) treatments, $\delta$$=$$-5$$\gamma$, $R$$\approx$$0.22$$\gamma$ ($I$$\approx$$20\,$mW/cm$^2$).}\label{fig5}
\end{figure}

\begin{figure*}
\centerline{\scalebox{0.9}{\includegraphics{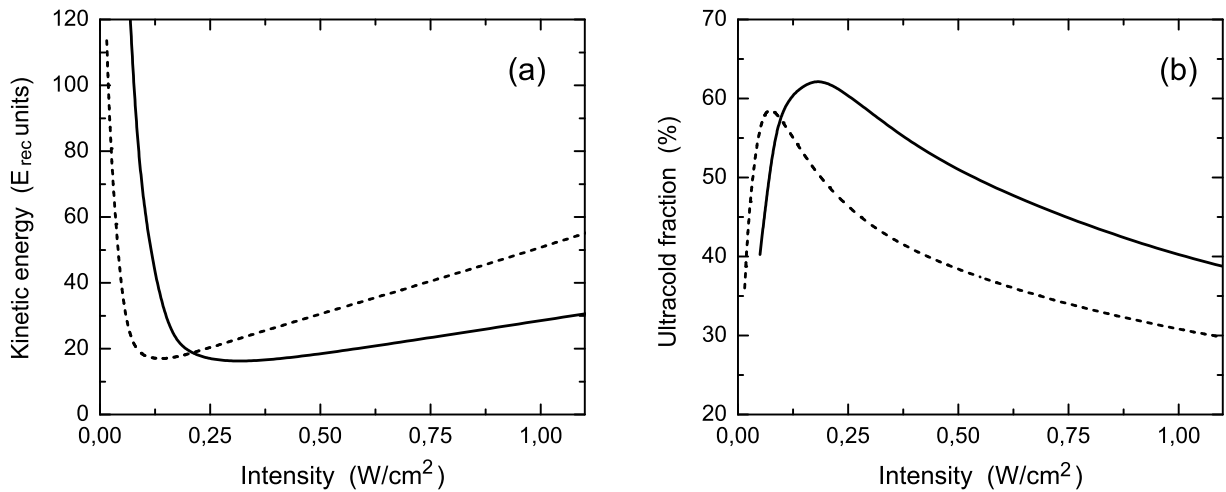}}}
\caption{(a) Average kinetic energy of an atom under {\it lin}$\perp${\it lin} light field configuration calculated on the basis of quantum treatment, (b) ultracold fraction of atoms in a cloud. Light field detunings are $\delta$$=$$-2$$\gamma$ (dashed) and $\delta$$=$$-5$$\gamma$ (solid).}\label{fig6}
\end{figure*}

\noindent Figure \ref{fig4}a shows average kinetic energy of an atom as the function of light field intensity, calculated on the basis of numerical solving the equation (\ref{DensityRotated2}). Analogical dependence is also presented, gained by applying semiclassical approach on the basis of Fokker-Planck equation (\ref{Semiclassics}). As it is seen from the figure the semiclassical approach (dashed line) gives the minimum kinetic energy of an atom at the level of $E_{min}$$\approx$$\,30$$E_{rec}$, what is several times smaller than the Doppler limit $E_D$$\approx$$\,87.5$$E_{rec}$. At the same time the quantum approach (solid line) shows the result for energy just a little bit smaller than the Doppler limit ($E_{min}$$\approx$$\,62$$E_{rec}$). Hence the quantum treatment of the problem shows that it is hardly possible to cool magnesium atoms in MOT down to desirable range of temperatures on the basis of transition $3^3$P$_2$$\to$$3^3$D$_3$. All this agrees with the experiments of research group from the University of Hannover \cite{MgDeep1,MgDeep2}. Effective temperature, corresponding to the minimum at the plot $E(I)$ for quantum-treatment result, is about 310~$\mu$K at frequency detuning $\delta$$=$$-5\gamma$$\approx$$-2\pi$$\times$$130\,$MHz and light field intensity $I$$\approx$$1100\,$mW/cm$^2$.

Beside the temperature of an atomic ensemble it is also important to know a profile of momentum distribution of atoms in a cloud. It may be found very useful, in particular, for realization of evaporative cooling stage for achieving ultralow temperatures ($\sim\,$1~$\mu$K). Let us consider a group of atom in the momentum space with $p$$\,\le$$\,3$$\,\hbar k$. Tentatively speaking we call this fraction as ``ultracold'' one. Figure~\ref{fig4}b shows number of atoms in the ultracold fraction $N_c$ as the function of light field intensity $I$. As it is seen from the figure there is a maximum in vicinity of 500~mW/cm$^2$. It should be noted that position of this optimum is not immediately the same as for the minimum of the dependence $E_{kin}(I)$. Figure~\ref{fig4}b demonstrates that about 40~\% of atoms can be concentrated in the ultracold fraction. For comparison analogical dependence is presented, calculated on the basis of semiclassical approach (dashed line), which lies noticeably higher than the former one. The dependencies $E_{kin}(I)$ and $N_c(I)$ lose to the semiclassical ones, because quantum treatment provides significantly different result for the momentum distribution in the vicinity of $p$$\approx$$0$. Indeed, Fig.~\ref{fig5} shows very sharp spike in the semiclassical case and a tiny peak as the result of quantum calculations.

\subsection{The second stage: Cooling in an optical molasses}
Fortunately, solution of the problem of deep laser cooling of magnesium atoms can be found by involving the second stage of sub-Doppler cooling with the help of one-dimensional optical molasses. The molasses is composed of two counterpropagating light waves with orthogonal linear polarizations ({\it lin}$\perp${\it lin} configuration).

In contrast to $\sigma^+$$\sigma^-$ configuration, in the case of {\it lin}$\perp${\it lin} field the total-field polarization transforms from linear to circular (and back) along z-axis (e.g., see \cite{Dalibard}). Consequently, there is no any rotating transformation of coordinate frame $K$ that would make density matrix independent of $z$-coordinate. Therefore, we must solve the set of equations (\ref{DensityBlocks})-(\ref{DensityBlocks4}) on matrix $\widehat{\rho}(z,q)$. We have solved the equation numerically on the basis of matrix continued fractions method. The details of the method can be found, for example, in \cite{Prudnikov} and we do not reproduce it here. Instead of that, we just present the numerical results.

\begin{figure}[t]
\centerline{\scalebox{0.8}{\includegraphics{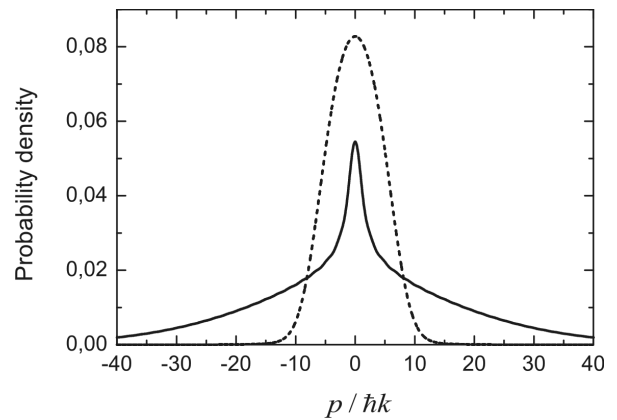}}}
\caption{Quantum calculations of momentum distributions at $\delta$$=$$-5$$\gamma$. Field strengths: $R$$\approx$$0.22$$\gamma$, $I$$\approx$$20\,$mW/cm$^2$ (solid) and $R$$\approx$$1.13$$\gamma$, $I$$\approx$$600\,$mW/cm$^2$ (dashed).}\label{fig7}
\end{figure}

Figure~\ref{fig6}a demonstrates much lower minimum kinetic energy than in the case of $\sigma^+$$\sigma^-$ field (see Fig.~\ref{fig4}a, solid line). In particular, the minimum corresponds to $E$$\approx$$\,16$$E_{rec}$ at $I$$\approx$$\,300$~mW/cm$^2$ ($T_{eff}$$\approx$$\,80\,$$\mu$K). Besides, as it is seen from Fig.~\ref{fig6}b, ultracold fraction of atoms under {\it lin}$\perp${\it lin} light field can be higher than in the case of $\sigma^+$$\sigma^-$ (compare with Fig.~\ref{fig4}b). Narrow structure in momentum profile in vicinity of $p$$\approx$$\,0$ becomes more visible than under $\sigma^+$$\sigma^-$ field (compare Fig.~\ref{fig7} and Fig.~\ref{fig5}). Therefore, the second cooling stage involving optical molasses can provide lower temperature as well as larger number of atoms in ultracold fraction (up to 60\%). After the second sub-Doppler cooling stage atoms may be loaded, for instance, to a dipole trap. At that the ``hot'' fraction of atoms (wide background at Fig.~\ref{fig7}) can be evaporated by proper choice of the light potential depth, saving only the ultracold fraction in a trap (with effective temperature $\sim1$$\,\mu$K). It should be noted that realization of the second sub-Doppler stage should eventually provide much more number of ultracold atoms in a dipole trap (or an optical lattice) after evaporation than without this stage.

\section{Conclusion}
In conclusion we would like to summarize main results of the work. We have suggested using the second sub-Doppler cooling stage to solve the problem of deep laser cooling of magnesium atoms. The first stage implies using of magneto-optical trap involving dipole transition between triplet states $3^3$P$_2$ and $3^3$D$_3$. In particular, this stage was used in the experiment of researches from Hannover University \cite{MgDeep2}. In spite of that the level $3^3$P$_2$ is degenerate and one could anticipate activation of effective sub-Doppler mechanism of cooling in polarization-gradient field \cite{Dalibard}, however, the conducted theoretical analysis has figured out the minimum temperature at the level of 310~$\mu$K, what is just a little bit lower than the estimate for Doppler limit of cooling ($T_D$$\approx\,$$425\,$$\mu$K). To reduce this value by several times we have proposed using the second laser-cooling stage with the help of optical molasses composed of two counterpropagating orthogonally linearly polarized waves ({\it lin}$\perp${\it lin} field configuration). In contrast to $\sigma^+$$\sigma^-$ field, applied in MOT, the optical molasses demonstrates much lower temperature (80~$\,\mu K$). At the same time, the minimum achievable temperature of laser cooling in the case of $^{24}$Mg still is noticeably higher than for some other atoms, where sub-Doppler mechanism provides much better results. Most likely, it is due to relatively large recoil energy of magnesium atom. For instance, for considered transition in magnesium recoil frequency $\omega_{rec}$$\approx\,$$2\pi$$\times$$53\,$kHz, while for $^{133}$Cs ($^2$S$_{1/2}$, F=4 $\to$ $^2$P$_{3/2}$, F=5) this frequency is significantly smaller ($\omega_{rec}$$\approx\,$$2\pi$$\times$$2\,$kHz), what allowed cooling cesium atoms down to 2.5~$\mu$K \cite{SalomonCs}.

Besides temperature (average kinetic energy) of ensemble of atoms we have also paid attention to the linear momentum distributions in the steady state. In particular, we have investigated the problem of increasing concentration of atoms in ultracold fraction (a region in momentum space in the vicinity of $p$$=$$0$). Conducted numerical calculations have revealed the optimum parameters of laser field for maximization of the ultracold fraction ($T_{eff}$$\sim$1~$\mu$K). This fraction can be easily localized in a optical trap, while the other fraction (``hot'' atoms) can be evaporated from the trap by proper choice of the optical depth. At that, it is the second stage of sub-Doppler cooling that can provide great increase of ultracold atomic number in comparison with the case with only the first stage realized (as in the experiments \cite{MgDeep1,MgDeep2}).

In our theoretical analysis quantum treatment with full account for the recoil effect has been exploited, i.e. we have not been limited by semiclassical or secular approximations as well as weak-field limit. It has allowed us studying kinetics of cold magnesium in a wide range of intensity and frequency detuning to determine the optimum parameters of laser field. Also we have compared data provided by quantum and semiclassical approaches. As the result of that comparison we can conclude that semiclassical approach in the case of transition $3^3$P$_2$$\to$$3^3$D$_3$ in $^{24}$Mg is not valid for adequate understanding the kinetics of ultracold magnesium atoms for a wide range of light-field parameters. Moreover, we can also conclude that for getting the adequate estimate of cooling parameters and understanding the problems in deep laser cooling of atoms it is quite necessary to treat the problems with the help of quantum approach.

At the end we should note that in spite of theoretical analysis has been done out of many widely used approximations, we have assumed the problem to be one-dimensional. However, light-field configuration used in a magneto-optical trap is always three-dimensional (three pairs of circularly polarized beams). Therefore, obviously, the results of such 1D analysis may differ from the real experiment with 3D field. For example, one can refer to the papers \cite{1Dvs3D1,1Dvs3D2} for getting the estimate of such kind of difference. In these papers calculations were done for 1D and 3D configurations by the example of simple transition $F_g$$=$$0$$\to$$F_e$$=$$1$ in limits of semiclassical and slow-atoms approximations. At the same time, an optical molasses (the second cooling stage suggested), which is of the most interest from the viewpoint of deep laser cooling, can be implemented in 3D as well as in 1D configuration. Three-dimensional optical molasses for various transition of the type $F_g$$=$$F$$\to$$F_e$$=$$F$$+$$1$ was investigated in ref. \cite{Castin3D} under weak-saturation approximation and with the help of adiabatic reduction of density matrix equations to ground state. Unfortunately that approximation gives good results not for wide range of parameters $\delta$ and $I$ that can have an interest from the laser cooling view of point (e.g., see the work \cite{Prudnikov}, where results of adiabatic approximation were compared with results of full quantum treatment). Three-dimensional quantum treatment with full account for the recoil effect and beyond the aforesaid approximations is quite difficult task that requires separate study.

\begin{acknowledgments}
The work was partially supported by the Ministry of Education and Science of the Russian Federation (gov. order no. 2014/139, project no. 825), Presidium of the Siberian Branch of the Russian Academy of Sciences, and by the grants of RFBR (nos. 15-02-06087, 15-32-20330, 14-02-00806, 14-02-00712, 14-02-00939) and Russian Presidential Grants (MK-4680.2014.2 and NSh-4096.2014.2).
\end{acknowledgments}

\end{document}